\documentclass[a4paper,twocolumn,pra,amsmath,amssymb]{revtex4}
\usepackage{graphicx}
\usepackage{epstopdf}
\usepackage{braket}

\newcommand{\Op}[1]{\boldsymbol{\mathsf{\hat{#1}}}}

\def\openone{\leavevmode\hbox{\small1\kern-3.3pt\normalsize1}}
\newcommand{\unity}{\openone}
\newcommand{\vphi}{\varphi}
\newcommand{\Tr}{\operatorname{tr}}
\newcommand{\tr}{\operatorname{tr}}
\newcommand{\eff}{\operatorname{eff}}
\newcommand{\diag}{\operatorname{diag}}
\newcommand{\lec}{\operatorname{LEC}}
\newcommand{\PE}{\operatorname{PE}}
\newcommand{\LI}{\operatorname{LI}}
\newcommand{\errPE}{\varepsilon_{\PE}}
\newcommand{\errLec}{\varepsilon_{\lec}}

\begin{document}

\title{Optimizing for an arbitrary perfect entangler. II.~Application}

\author{Michael H. Goerz}
\affiliation{Theoretische Physik, Universit\"{a}t Kassel,
  Heinrich-Plett-Str. 40, D-34132 Kassel, Germany}

\author{Giulia Gualdi}
\affiliation{Theoretische Physik, Universit\"{a}t Kassel,
  Heinrich-Plett-Str. 40, D-34132 Kassel, Germany}

\author{Daniel M. Reich}
\affiliation{Theoretische Physik, Universit\"{a}t Kassel,
  Heinrich-Plett-Str. 40, D-34132 Kassel, Germany}

\author{Christiane P. Koch}
\affiliation{Theoretische Physik, Universit\"{a}t Kassel,
  Heinrich-Plett-Str. 40, D-34132 Kassel, Germany}

\author{Felix Motzoi}
\affiliation{Department of Chemistry, 
University of California, Berkeley, California 94720, USA}

\author{K. Birgitta Whaley}
\affiliation{Department of Chemistry, 
University of California, Berkeley, California 94720, USA}

\author{Ji\v{r}\'i Vala}
\affiliation{Department of Mathematical Physics, National University
  of Ireland, Maynooth, Ireland}
\affiliation{School of Theoretical Physics, Dublin Institute for
  Advanced Studies, 10 Burlington Rd., Dublin, Ireland}

\author{Matthias M. M\"uller}
\affiliation{Center for Integrated Quantum Science and Technology,
  Institute for Complex Quantum Systems, Universit\"at Ulm, D-89069 Ulm, Germany}

\author{Simone Montangero}
\affiliation{Center for Integrated Quantum Science and Technology,
  Institute for Complex Quantum Systems, Universit\"at Ulm, D-89069 Ulm, Germany}

\author{Tommaso Calarco}
\affiliation{Center for Integrated Quantum Science and Technology,
  Institute for Complex Quantum Systems, Universit\"at Ulm, D-89069 Ulm, Germany}

\date{\today}

\begin{abstract}
  The difficulty of an optimization task in quantum information
  science depends on the proper mathematical expression of the
  physical target. Here we demonstrate the power of optimization
  functionals targeting an arbitrary perfect two-qubit entangler,
  creating a maximally-entangled state out of some initial product
  state. For two quantum information platforms of current interest,
  nitrogen vacancy centers in diamond and superconducting Josephson
  junctions, we show that
  an arbitrary perfect entangler can be reached faster and
  with higher fidelity than specific two-qubit gates or local
  equivalence classes of two-qubit gates. Our results are obtained
  with two independent optimization approaches, underlining the
  crucial role of the optimization target.
\end{abstract}


\maketitle

\section{Introduction}
\label{sec:intro}

Optimal control theory \cite{brif2010} is a versatile tool to tackle tasks in quantum
information science, allowing to reach very high fidelities in state
preparation~\cite{RojanPRA14,Hoyer2014,Montangero2007}
and manipulation~\cite{PalaoPRL02,ToSHJPB11,MuellerPRA11,Muller2014,VanFrank2014} for complex quantum
systems. Very recently the field of application of optimal control has also
been extended also to many body quantum systems~\cite{Caneva2011,Caneva2011a, Caneva2013, Doria2011,MuellerPRA13}.
Optimal control represents a mathematical toolbox which processes its input, the
desired physical target and constraints as well as the quantum
system's equation of motion, to yield external controls that drive the
dynamics towards the target as best possible. The optimized controls
thus depend crucially on a proper mathematical formulation of the
physical ingredients. For example, it is possible to obtain controls
that are robust with respect to experimentally unavoidable
fluctuations by accounting for these fluctuations in the
optimization~\cite{kobzar2004nmr,kobzar2008nmr,GoerzPRA14}.
Similarly, when treating the quantum
system as open, one can explore the limits on fidelity imposed by
decoherence~\cite{ToSHJPB11,FloetherNJP12,GormanPRA12,GoerzNJP14,Mukherjee2013} or
identify control mechanisms that rely on the coupling to the
environment~\cite{RebentrostPRL09,SchmidtPRL11,Reich14}.
Optimal control also allows to develop time-optimal strategies,
resulting in protocols that perform transformations at the fastest possible pace -- the so
called Quantum Speed Limit (QSL) --
compatible with energy and information constraints~\cite{CampoPRL2013,DeffnerPRL2013,LevitinPRL2009,MargolusPD1998,BhattacharyyaJPA1983,Lloyd2014}.

While fluctuations and coupling to the environment enter the equation
of motion, the optimization goal and additional constraints are
expressed in the optimization functional. One can target a
state-to-state transition~\cite{SomloiCP93}, a quantum
gate~\cite{PalaoPRL02}, or a certain class of quantum
gates~\cite{MuellerPRA11}. It is also possible to minimize the system
energy~\cite{Caneva2011a},
maximize entanglement~\cite{PlatzerPRL10},
or prescribe a desired time evolution~\cite{SerbanPRA05}, as well as targeting an unknown
stable and maximally entangled states~\cite{Caneva2012}.
Typical constraints include finite pulse energy~\cite{SomloiCP93} and
bandwidth~\cite{ReichJMO14,PalaoPRA13} or smoothness of the
control~\cite{HohenesterPRA07}. Constraints naturally limit the
resources available for control and thus restrict the
search. This does not only slow down convergence of the optimization,
but may also prevent reaching the target with sufficient fidelity
altogether~\cite{Rabitz2004,MooreJCP12}.

Similarly, formulating the optimization target in an overly specific
way may unnecessarily restrict the flexibility of optimization. For
example, in the circuit model of quantum computing, the capability to
implement an entangling two-qubit gate is
required~\cite{NielsenChuang}. This may be the controlled NOT gate but
any gate within the local equivalence class of CNOT, i.e., all gates
that differ from CNOT only by single qubit operations, 
will work equally well~\cite{ZhangPRA03}. However,
the time evolution of gates in the same local equivalence class is
generated by Hamiltonians which may be very
different. For example, diagonal Hamiltonians are sufficient to
generate a controlled phasegate which is locally equivalent to CNOT,
whereas CNOT itself requires non-diagonal entangling terms. Thus,
optimization for the CNOT gate does not have any effect if the
Hamiltonian is diagonal but targeting an arbitrary gate in the local
equivalence class of CNOT with the same Hamiltonian is
effective~\cite{MuellerPRA11}. The corresponding optimization
functional utilizes the geometric theory of non-local two-qubit
operations~\cite{ZhangPRA03}.

For other applications in quantum information science, the
optimization target can be formulated even more generally than the
functional for a gate
within a certain local equivalence class. For example, the capability
to implement an arbitrary perfect entangler (PE) is sufficient
in quantum communication. 
Perfect entanglers make up a large part of all entangling two-qubit
operations. Therefore, extending the optimization functional targeting
a specific local equivalence class to comprise all perfect entanglers
holds the promise of a significantly more flexible and easier search.
This flexibility is crucial when very high fidelities are required or
when  optimal control theory is utilized to
identify fundamental limits for control in a numerical search.

In the preceding paper~\cite{Watts}, two variants of
an optimization functional targeting all two-qubit perfect
entanglers have been developed.
Here, we apply these functionals to two quantum
information platforms that currently enjoy great
popularity, nitrogen vacancy (NV) centers in diamond and
superconducting Josephson junctions,
due to their promise of control and scalability.
%
%
In particular, both are candidates for use in hybrid systems which
shall provide the interconnect between information processing and
communication platforms.

The two variants of the perfect entanglers functional
differ in the representation of the
two-qubit non-locality, utilizing either the coefficients $c_1$,
$c_2$, $c_3$ of $\sigma_x\otimes\sigma_x$, $\sigma_y\otimes\sigma_y$,
$\sigma_z\otimes\sigma_z$ in the canonical parametrization of
two-qubit gates or the so-called local
invariants~\cite{ZhangPRA03}. The latter offer the advantage that they
can be calculated from the time evolution in a closed analytical
formula. This is a prerequisite for optimization algorithms using
gradient information~\cite{MuellerPRA11,ReichJCP12}. Due to the
non-linear relation between the coefficients $c_1$, $c_2$, $c_3$
and the local invariants, the
topology underlying the two optimization functionals is rather
different~\cite{Watts}. While the $c$-space (the space spanned by the
coefficients $c_1$, $c_2$, $c_3$; we refer to this also as the Weyl chamber due
to the symmetries of the $c$-space -- see 
Ref.~\onlinecite{ZhangPRA03})
variant is expected to
provide for a more direct approach towards the target, gradient
algorithms can use more information of a given
control landscape. In order to investigate whether the
topology underlying the optimization functional influences the final
fidelities, we employ
two different optimization methods, Chopped Random Basis (CRAB)
optimization, which combines a gradient-free search with a randomized
parametrization of the control~\cite{DoriaPRL11,CanevaPRA11}, and
Krotov's method, which utilizes gradient information
~\cite{ReichJCP12}.

Our paper is organized as follows. We first present the models for our
applications, NV centers in diamond and superconducting charge and
transmon qubits, in Sec.~\ref{sec:models}. Section~\ref{sec:CRAB}
reviews the CRAB optimization algorithm which utilizes the perfect
entanglers functional based on the coefficients $c_1$, $c_2$,
$c_3$. Krotov's method with the corresponding perfect entanglers
functional written in terms of the local invariants is presented in
Sec.~\ref{sec:Krotov}. Our numerical results are discussed in
Sec.~\ref{sec:appl} for the three models, and we conclude in
Sec.~\ref{sec:concl}.

\section{Models}
\label{sec:models}

An optimization functional that allows for a very flexible search is
only useful when the system dynamics is sufficiently complex to
explore different areas of the search space.
In paper I~\cite{Watts}, the two-qubit Hamiltonian
\begin{eqnarray}
  \Op{H}
  & = &
      \sum_{\alpha=1,2} \frac{\omega_{\alpha}}{2} \Op{\sigma}_z^{(\alpha)}
      + u_1(t) \left( \Op{\sigma}_x^{(1)} + \lambda \Op{\sigma}_x^{(2)} \right)
  + \nonumber \\ & &
      + u_2(t) \left(
          \Op{\sigma}_x^{(1)} \Op{\sigma}_x^{(2)}
         +\Op{\sigma}_y^{(1)} \Op{\sigma}_y^{(2)}
        \right)
  \label{eq:fullH}
\end{eqnarray}
with $\Op{\sigma}_{i}^{\left(j\right)}$  the $i$th Pauli operator
acting on the $j$th qubit and $u\left(t\right)$ the control field,
was shown to allow for a non-trivial search in the Weyl chamber.
The Weyl chamber is the geometric space spanned by the non-local coefficients
$c_1$, $c_2$, $c_3$, taking into account reflection symmetries, see
Ref.~\onlinecite{ZhangPRA03}.
Here, we extend the discussion to specific physical examples, starting
with an NV center in diamond, followed by superconducting transmon and
charge qubits. The Hamiltonians of the latter two can be related to
Eq.~\eqref{eq:fullH}, although for realistic parameters, additional
levels, beyond the logical two-qubit space, have
to be taken into account.


\subsection{NV$+^{13}$C center in diamond}
\label{subsec:model:NV}

\begin{figure}[tb] 
  \centering
  \includegraphics{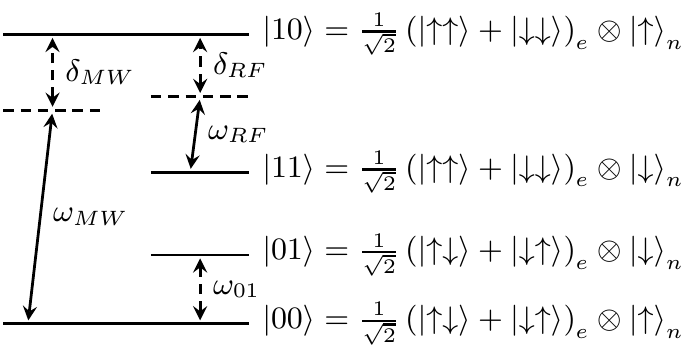}
  \caption{Level structure for NV$+^{13}$C center. The electronic
    states $|0\rangle_e$ and $|1\rangle_e$ correspond to the ground
    $m_s=0$ and degenerate $m_s = \pm 1$ sublevels of the $S=1$
    triplet state of the two unpaired  electrons of the NV center (see
    Ref.~\onlinecite{nizovtsev2005quantum}). The energy levels are not
    drawn to scale.} 
  \label{fig:nvlevels}
\end{figure}
For an NV center in diamond, we employ the NV+$^{13}$C model of
Refs.~\cite{nizovtsev2005quantum,SaidPRA2009} for the ground $^3$A state of the NV center coupled to a $^{13}$C nuclear spin.  This comprises the four states
$|00\rangle=|0\rangle_e|0\rangle_n$, $|01\rangle=|0\rangle_e|1\rangle_n$,
$|10\rangle=|1\rangle_e|0\rangle_n$, and
$|11\rangle=|1\rangle_e|1\rangle_n$,
where the electronic states $|0\rangle_e$ and $|1\rangle_e$ correspond to the ground $m_s=0$ and degenerate $m_s = \pm 1$ sublevels of the $S=1$ triplet state of the two unpaired  electrons of the NV center (see Figure~\ref{fig:nvlevels}).
Transitions between these levels are driven by a radio
frequency field,
\begin{equation}
  \epsilon_{RF}(t)
  = \Omega_{RF}(t) \cos\left[ (\omega_{RF} - \delta_{RF}(t)) \,t\right]\,,
\end{equation}
and a microwave field,
\begin{equation}
  \epsilon_{MW}(t)
  = \Omega_{MW}(t) \cos\left[ (\omega_{MW} - \delta_{MW}(t)) \,t\right]\,,
\end{equation}
as depicted in Fig.~\ref{fig:nvlevels}. In the interaction picture
($\omega_{01} = 0$) under the rotating wave approximation, the
corresponding Hamiltonian reads
\begin{equation}
  \label{eq:H_NV}
  \Op{H}_{NV} =
  \begin{pmatrix}
    0 & 0 & \frac{\Omega_{MW}(t)}{2} & 0 \\
    0 & 0 & 0 & 0 \\
    \frac{\Omega_{MW}(t)}{2} & 0 & \delta_{MW}(t) & \frac{\Omega_{RF}(t)}{2}\\
    0 & 0 & \frac{\Omega_{RF}(t)}{2} & \Delta(t)
  \end{pmatrix}\,,
\end{equation}
with $\Delta(t) = \delta_{MW}(t)-\delta_{RF}(t)$.
If we tune the radiation fields onto resonance,
$\delta_{MW}(t) \equiv \delta_{RF}(t) \equiv 0$,
only two driving terms are retained in Eq.~\eqref{eq:H_NV}, i.e.,
$\Op{H}^{\Delta=0}_{NV} = \Op{H}_{MW} + \Op{H}_{RF}$ with
\begin{eqnarray}
 \label{eq:H_MW}
 \Op{H}_{MW}
  &=&\frac{\Omega_{MW}(t)}{4}
   \Op{\sigma}_x^{(1)}\otimes(\unity+\Op{\sigma}_z^{(2)})\,, \\
 \label{eq:H_RF}
 \Op{H}_{RF}
 &=&\frac{\Omega_{RF}(t)}{4}
  (\unity-\Op{\sigma}_z)^{(1)}\otimes\Op{\sigma}_x^{(2)}\,.
\end{eqnarray}
For a gate duration of $T = 5\,\mathrm{\mu\,s}$, typical amplitudes for the
optimized pulses are on the order of $\Omega_{MW}/2\pi = 50\,$MHz and
$\Omega_{RF}/2\pi = 100\,$kHz.

The interaction described by Eqs.~\eqref{eq:H_MW} and~\eqref{eq:H_RF}
is of different
form than Eq.~\eqref{eq:fullH}, and thus we have to analyze controllability
separately. This is straightforward: Labelling
$\Op{A}_1=\Op{\sigma}_x^{(1)}\otimes (\unity+\Op{\sigma}_z)^{(2)}$,
$\Op{A}_2=(\unity-\Op{\sigma}_z)^{(1)}\otimes \Op{\sigma}_x^{(2)}$,
their commutator yields
\begin{equation}
\Op{A}_3
 =[\Op{A}_1,\Op{A}_2]
 = -\Op{\sigma}_x\Op{\sigma}_y
   -\Op{\sigma}_y\Op{\sigma}_x\,.
\end{equation}
The nested commutators read
\begin{equation}
 [\Op{A}_1,\Op{A}_3]= -\Op{A}_2, \quad
 [\Op{A}_2,\Op{A}_3]=\Op{A}_1\,.
\end{equation}
Therefore, the Lie algebra is closed under just the three operators
$\Op{A}_1$, $\Op{A}_2$, $\Op{A}_3$.
These operators are not linearly independent
(up to local transformations) and correspond to only two dimensions of
the Weyl chamber, specifically the ground plane.

We also consider the case of a non-zero detuning $\Delta(t)$ in
Eq.~\eqref{eq:H_NV}. This results in a third control Hamiltonian,
\begin{eqnarray}
  \label{eq:DeltaControl}
  \Op{H}_\Delta
  & = &
  \diag(0, 0, 0, \Delta(t))
  \nonumber \\
  &=& \frac{\Delta(t)}{4} \left(
      \unity - \Op{\sigma}_z^{(1)} - \Op{\sigma}_z^{(2)}
      + \Op{\sigma}_{z}\Op{\sigma}_z
    \right)\,.
\end{eqnarray}
$\Delta(t)$ is on the order of 1$\,$MHz.
This additional terms provides the missing commutators necessary to reach every
point in the Weyl chamber. Both cases, $\Op{H}^{\Delta=0}_{NV}$ and
$\Op{H}_\Delta$, allow for a non-trivial search in the
Weyl chamber, making it a suitable candidate for optimization
employing the perfect entanglers functional.

\subsection{Charge qubits with Josephson junction coupling}

\label{subsec:model:charge}
\begin{figure}[tb] 
  \centering
  \includegraphics{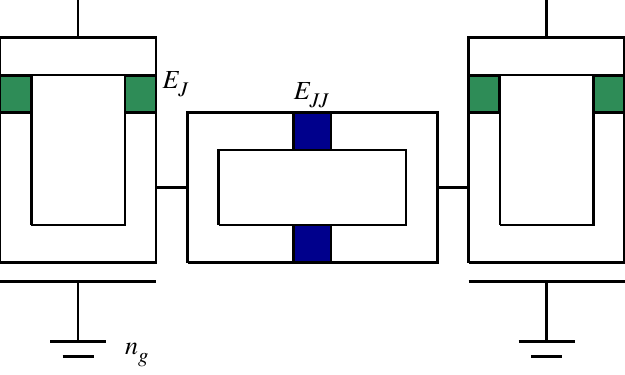}
  \caption{(Color online) Setup of the Josephson charge qubits (green, left and right) coupled by
  a Josepshon junction (blue, middle). The local charge levels are driven by $E_J$.
  The interaction is driven by $E_{JJ}$.
  }
  \label{fig:JCQ-sketch}
\end{figure}
As an example relating more directly to Eq.~\eqref{eq:fullH}, we consider two
superconducting charge qubits coupled via a Josephson junction
\cite{MontangeroPRL2007}, as depicted in Fig.~\ref{fig:JCQ-sketch}.
The local Hamiltonian reads
\begin{eqnarray}
  \label{eq:H_charge_loc}
  \Op{H}^{\text{loc}}_C
  & = &
  \sum_{i=1,2} \sum_{n_i}
  \Bigg[
    E_C \left(n_i -n_g^{(i)}\right)^2 \Ket{n_i}\!\Bra{n_i}
    \nonumber \\ & &
    - \frac{E_J^{(i)}(t)}{2}
     \bigg(
      \Ket{n_i}\!\Bra{n_{i} + 1} + \Ket{n_{i}+1}\!\Bra{n_i}
    \bigg)
  \Bigg]\,.
\end{eqnarray}
We can control the Josephson coupling $E_j^{(1)}(t)=E_j^{(2)}(t)$, while the
charging energy $E_C$ and the offset charge $n_g^{(i)}$ are fixed.
In order to make the connection to Eq.~\eqref{eq:fullH} explicit, we
truncate each anharmonic ladder to two levels,
\begin{equation}
  \Op{H}^{\text{loc}}_{C,2l}
  = \sum_{i=1}^{2} \left[
      E_C \left(n_g - \frac{1}{2}\right) \Op{\sigma}_z^{(i)}
      - \frac{E_J(t)}{2} \Op{\sigma}_{x}^{(i)} \right]\,,
\end{equation}
omitting terms proportional to the identity. As was shown in
Paper~I~\cite{Watts}, only
a two-dimensional subsection of the Weyl chamber can be reached at the
degeneracy point, i.e.  $n_g = \frac{1}{2}$. Here, as we are more interested in
the control problem than in its possible experimental realization and thus its
robustness against noise, we set $n_g = 1$, lifting the degeneracy between the
qubit levels and ensuring full controllability.

The interaction is described by
\begin{equation}
  \label{eq:H_charge_int}
  \Op{H}_{JJ}
  = \frac{E_{JJ}(t)}{2}\sum_{n_1,n_2} \left(
    \Ket{n_1,n_2+1}\!\Bra{n_1+1,n_2} + h.c.
    \right)\,,
\end{equation}
or
\begin{equation}
  \Op{H}_{JJ,2l}
  =
  \frac{E_{JJ}(t)}{4} \left(
      \Op{\sigma}_{x}^{(1)} \Op{\sigma}_{x}^{(2)}
     +\Op{\sigma}_{y}^{(1)} \Op{\sigma}_{y}^{(2)}
  \right)\,,
\label{spinB}
\end{equation}
if truncated to two levels.
Typically, $E_C$ takes values between 20$\,$GHz and 200$\,$GHz, and
$E_J/E_C$ is between
$0.03$ and $0.5$. Both single and two-qubit gates can be implemented on
a picosecond time scale \cite{NakamuraN1999, MakhlinRMP2001, YamamotoN2003}.
The two-level truncation of the Hamiltonian is generally only accurate near the
charge degeneracy point. For the parameters considered here, higher
levels have to be taken into account.

\subsection{Two transmon qubits interacting via a cavity}
\label{subsec:model:transmon}

The transmon qubit \cite{JKochPRA07} is closely related to the charge
qubit, but operated in a different parameter regime,  $E_J
\gg E_C$. This makes them significantly less anharmonic than typical
charge qubits, but more robust with respect to charge noise.
The coupling between two transmons is implemented via a shared transmission
line resonator (``cavity''). The  energy of
each transmon qubit transition is denoted by $\omega_1$, $\omega_2$
for the first (``left'') and second (``right'') transmon, respectively.
Higher levels are given as a Duffing oscillator with anharmonicity
$\alpha_1$, $\alpha_2$. Each qubit couples to the cavity with coupling
strength $g_1$ , $g_2$.
In the dispersive limit $|\omega_i -\omega_r| \gg |g_{i}|$ ($i=1,2$) with
$\omega_r$ the cavity frequency, the cavity can be eliminated and an effective
two-transmon Hamiltonian is obtained. The coupling between each
transmon and the cavity turns into an effective qubit-qubit coupling,
\begin{equation}
J^{\eff}
\approx
    \frac{g_1 g_2}{(\omega_1-\omega_r)}
  + \frac{g_1 g_2}{(\omega_2-\omega_r)}\,.
\end{equation}
In most current setups, $J^{\eff} \ll |\omega_2 - \omega_1|$,
and the two-transmon Hamiltonian can be approximated as \cite{PolettoPRL2012}
\begin{eqnarray}
  \Op{H}_{T}
  & \approx &
    \sum_{i=1,2} \left(
        \left( \omega_i + \frac{\alpha_i}{2}\right)
        \Op{b}_i^{\dagger} \Op{b}_i
        - \frac{\alpha_i}{2} \left( \Op{b}_i^{\dagger} \Op{b}_i \right)^2
    \right)
    \nonumber \\ & &
    + J^{\eff} \left( \Op{b}_1^\dagger \Op{b}_2
                  + \Op{b}_1 \Op{b}_2^\dagger
            \right)
            \nonumber \\ & &
            + \Omega(t) \left( \Op{b}_1 + \Op{b}_1^\dagger
              + \lambda \Op{b}_2 + \lambda \Op{b}_2^\dagger \right)\,,
\label{eq:H_transmon}
\end{eqnarray}
where $\Omega(t)$ is the driving field that couples to the cavity. Typical
parameters are listed in Table~{\ref{tab:transmon_parameters}}.
\begin{table}[tb]
  \centering
  \begin{tabular}{llrl} \hline\hline
  left qubit frequency           &  $\omega_1$  & 4.380 &GHz \\
  right qubit frequency          &  $\omega_2$  & 4.614 &GHz \\
  left qubit anharmonicity       &  $\alpha_1$  & -210  &MHz \\
  right qubit anharmonicity      &  $\alpha_1$  & -215  &MHz \\
  effective qubit-qubit coupling &  $J^{\eff}$  & -3.0  &MHz \\
  relative coupling strength     &  $\lambda$   & 1.03  &~\\
  \hline\hline
  \end{tabular}
  \caption{Parameters for the transmon Hamiltonian Eq.~\eqref{eq:H_transmon}}.
  \label{tab:transmon_parameters}
\end{table}
A Hamiltonian of the form of Eq.~\eqref{eq:fullH} is obtained by truncating the
higher levels,
\begin{eqnarray}
  \Op{H}_{T,2l}
  & = &
    \frac{\omega_i}{2} {\sigma^{(i)}_z}
    + 2 J^{\eff} \left( \Op{\sigma}_{x}^{(1)} \Op{\sigma}_{x}^{(2)}
                       +\Op{\sigma}_{y}^{(1)} \Op{\sigma}_{y}^{(2)} \right)
  + \nonumber \\ & &
    + \Omega(t) \left( \Op{\sigma}_x^{(1)} + \lambda
      \Op{\sigma}_x^{(2)}\right)\,.
\label{spinC}
\end{eqnarray}

Finally, notice that in systems B and C, the assumption that only two qubit
levels are populated leading to Eqs.~\eqref{spinB} and~\eqref{spinC} might not
hold in general, however, we present it to make the connection to
Eq.~\eqref{eq:fullH} and the controllability analysis of Paper~I~\cite{Watts}
explicit.

\section{Optimal control using CRAB}
\label{sec:CRAB}

\subsection{Perfect entangler functional in $c$-space}

A common choice for a fidelity that quantifies
whether an obtained gate $U$ corresponds to the target gate $V$ is
\cite{PalaoPRA03}
\begin{equation}
  F_{\text{sm}} =\frac{1}{4} \left| \tr \left[V^\dagger U\right] \right|^2\,.
\end{equation}
In the context of the geometric theory for two-qubit gates, reviewed
in paper~I~\cite{Watts}, where arbitrary local transformations are allowed,
the generalization of this fidelity can be expressed through the difference of
the Weyl chamber coordinates of the obtained gate $U$ and the target point,
\begin{equation}
 F_{\lec}(U)=\cos\frac{\Delta c_1}{2}\cos\frac{\Delta c_2}{2}\cos\frac{\Delta
    c_3}{2}\,.
\end{equation}
This fidelity can also be used as a functional for optimizing towards gates of a given
local equivalence class as shown in the preceding paper.
Building on the local equivalence class functional, a functional $F_{\PE}$ for
the optimization of an arbitrary perfect entangler can be derived.
In the Weyl chamber, the perfect entanglers form a polyhedron confined by the
three planes
\begin{subequations}
\begin{eqnarray}
 c_1+c_2&=&\pi/2\,,\\
 c_2+c_3&=&\pi/2\,\quad\text{and}\\
 c_1-c_2&=&\pi/2\,.
\end{eqnarray}
\end{subequations}
These planes divide the Weyl chamber into the polyhedron of perfect entanglers
and three corners of non-perfect entanglers. Within the perfect entangler
polyhedron, the functional is defined to take the value $F_{\PE}
\equiv 1$. Outside of the polyhedron, the value of $F_{\PE}$ depends
on the region  of the Weyl chamber the gate is in,
\begin{eqnarray}
 F_{\PE}(U)=\begin{cases}
\cos^2\frac{c_{U,1}+c_{U,2}-\frac{\pi}{2}}{4}\,,\qquad c_1+c_2\leq\frac{\pi}{2}\\
\cos^2\frac{c_{U,2}+c_{U,3}-\frac{\pi}{2}}{4}\,,\qquad c_2+c_3\geq\frac{\pi}{2}\\
\cos^2\frac{c_{U,1}-c_{U,2}-\frac{\pi}{2}}{4}\,,\qquad c_1-c_2\geq\frac{\pi}{2}\\
1\qquad\text{otherwise (inside polyhedron).}
          \end{cases}
\end{eqnarray}
As shown in the preceding paper, both $F_{\lec}$ and $F_{\PE}$ are not only optimization
functionals but have also a nonlocal fidelity interpretation.

Generally, the logical two-qubit subspace is embedded in a larger Hilbert space,
such that while the dynamics in the total Hilbert space are unitary, the
dynamics in the subspace may not be. In this case,
a closest unitary $U$ can be derived from the non-unitary (projected) gate
$\tilde{U}$: If $\tilde{U}$ has the singular value decomposition
$\tilde{U} = V \Sigma W^{\dagger}$, then the unitary that fulfills
$U = \arg \min_{u} \Vert u-\tilde{U} \Vert$ is given by $U = V W^\dagger$.
The local equivalence-class and perfect entangler fidelities then become
\begin{eqnarray}
F_{\lec}(\tilde U) = F_{\lec}(U) -||\tilde U -U||\,,\label{eq:FUtilde-lec}\\
F_{\PE}(\tilde U) = F_{\PE}(U) -||\tilde U -U||\,.\label{eq:FUtilde-pE}
\end{eqnarray}
These fidelities can be directly used as optimization functionals. The
optimization target is then to find $c_i$ in such a way that
Eqs.~\eqref{eq:FUtilde-lec} and~\eqref{eq:FUtilde-pE} are maximized.

\subsection{CRAB algorithm}

The chopped random basis (CRAB)
algorithm~\cite{DoriaPRL11,CanevaPRA11}  is an optimal control tool
that allows to optimize quantum operations in cases where it is either
not possible or impractical to calculate gradients of the optimization
functional.
In the present context, it is mathematically unfeasible to calculate
gradients of $F_{\lec}$ and $F_{\PE}$ as given in
Eqs.~\eqref{eq:FUtilde-lec}, \eqref{eq:FUtilde-pE} with respect to the
states (as needed for the Krotov update formula in
Sec~\ref{sec:Krotov} below),  since the functionals depends on the
states in a highly non-trivial way.

The central idea of CRAB is the expansion of the control function into
a truncated basis using random basis
functions~\cite{DoriaPRL11,CanevaPRA11},
\begin{equation}
 u(t)=\sum_{i=1}^n c_i f_i(t)\,,
\end{equation}
where the set of $f_i$ form the truncated basis.
We choose $f_i(t)=\sin (\omega_i t)$
with random $\omega_i\in [\frac{2\pi}{T}(i-0.5),\frac{2\pi}{T}(i+0.5)]$.
The coefficients $c_i$ are optimized by a direct search
algorithm, Nelder-Mead downhill simplex in our case.

\section{Optimal control using Krotov's method}
\label{sec:Krotov}

\subsection{Perfect entangler functional in $g$-space}
\label{subsec:imp_Krotov}

For optimal control approaches utilizing
gradient information, the capability to take
the derivative of the optimization functional
with respect to the unitary $U$, or equivalently, with respect to the
time-evolved basis states, is required. This is not possible for
the functionals in Eqs.~\eqref{eq:FUtilde-lec} and~\eqref{eq:FUtilde-pE}.
We therefore use an equivalent functional, based not on the Weyl
chamber coordinates $c_1$, $c_2$, $c_3$, but on
the local invariants $g_1$, $g_2$, $g_3$~\cite{Watts}.

An appropriate functional to optimize towards gates in a local
equivalence class is given by~\cite{MuellerPRA11}
\begin{equation}
  \label{eq:J_T_li_g}
  J_{\LI}(U) = (\Delta g_1)^2 + (\Delta g_2)^2 + (\Delta g_3)^2\,,
\end{equation}
where $\Delta g_i$ is the Euclidean distance between local invariant $g_i$ of
the obtained unitary $U$ and that of the optimal gate $O$.
For the perfect entanglers, the functional becomes~\cite{Watts}
\begin{equation}
  \label{eq:J_T_pe_g}
  \mathcal{D}(U)      =  g_3 \sqrt{g_1^{2} + g_2^{2}} - g_1\,.
\end{equation}
Both of these functionals take the value zero if the goal is reached.
They are thus distance measures, as opposed to the fidelities in
Eqs.~\eqref{eq:FUtilde-lec}, \eqref{eq:FUtilde-pE}, and they are not
restricted to lie within the range $[0,1]$.

Again, non-unitarity due to projection onto the logical
subspace must be taken into account. However, the
expression $||\tilde U -U||$, similarly to the functionals $F_{\lec}$,
$F_{\PE}$ used in Sec~\ref{sec:CRAB}, cannot easily be
differentiated. As an alternative, we minimize the loss of population
$\Tr\left[ \tilde U^\dagger \tilde U \right]/4$
from the logical subspace,
\begin{eqnarray}
  J_{\LI}(\tilde U)
  &=& w J_{\LI}(U) + (w-1)\left(1 - \frac{1}{4}
    \Tr\left[ \tilde U^\dagger \tilde U \right]\right)
  \label{eq:J_LI_tilde}\\
  \mathcal{D}(\tilde U)
  &=& w \mathcal{D}(U) + (w-1) \left(1 - \frac{1}{4}
    \Tr\left[ \tilde U^\dagger \tilde U \right]\right)\,.
  \label{eq:J_pe_tilde}
\end{eqnarray}
In Eqs.~\eqref{eq:J_LI_tilde}, \eqref{eq:J_pe_tilde},
the factor $w \in [0,1]$ is used to weight the relative importance of
Weyl-chamber optimization and unitarity. It can adaptively be
changed during the optimization in order to improve convergence.

\subsection{Krotov's Method}
\label{subsec:KrotovAlgo}

In Krotov's method, the total functional $J$ must include a
control-dependent running cost in order to derive an update
equation. $J$ takes the form
\begin{equation}
  \label{eq:J_krotov}
  J =
  J_T\left[\{\vphi_k(T)\}\right]
  + \int_0^T   \frac{\lambda_a}{S(t)}
    \left[ u(t) - u_\mathrm{ref}(t)\right]^2  \;dt \,,
\end{equation}
where $J_T$ is a final-time functional, e.g.\
Eq.~\eqref{eq:J_LI_tilde} or~\eqref{eq:J_pe_tilde}. The second term is
a constraint on the optimized control field $u(t)$.
Taking the reference field $u_{\mathrm{ref}}(t)$ to be the field from
the previous iteration ensures that close to the optimum the
functional is improved only due to changes in the actual target
$J_T$~\cite{PalaoPRA03}.

A comprehensive description of Krotov's method for quantum control
problems is found in Ref.~\cite{ReichJCP12}. Here, we state the control
equations for a final-time functional $J_T$ that depends higher than
quadratically on the states, linear coupling to the control and linear equations
of motion. In this case, the  update equation for the control at the $i+1$st
iterative step,
$u^{(i+1)}(t)$,  is given by
\begin{widetext}
\begin{equation}
\label{eq:newu}
  u^{(i+1)}(t)
  =
  u_\mathrm{ref}(t) +
  \frac{S(t)}{\lambda}
  \mathfrak{Im} \bigg\{\sum_{k=1}^4 \left\langle \chi_k^{(i)}(t)\Bigg|
      \frac{\partial \Op{H}}{\partial u}
      \bigg|_{u^{(i+1)}}
      \Bigg| \vphi_k^{(i+1)}(t)\right\rangle
   +\frac{1}{2}  \sigma(t)\sum_{k=1}^4
    \left\langle \Delta\vphi_k(t) \Bigg|
      \frac{\partial\Op{H}}{\partial u}\bigg|_{u^{(i+1)},}
      \Bigg|\vphi_k^{(i+1)}(t)\right\rangle \bigg\}
\end{equation}
\end{widetext}
with
$\Ket{\Delta\vphi_k(t)}=\Ket{\vphi_k^{(i+1)}(t)}-\Ket{\vphi_k^{(i)}(t)}$
representing the  change in state $\Ket{\vphi_k(t)}$.
In Eq.~\eqref{eq:newu}, $S(t)$ is a shape function to smoothly
switch the control on and off, and $\lambda$ is a parameter that
determines the step size of the change in the control. The scalar
function $\sigma(t)$ is constructed to ensure monotonic
convergence. For final-time functionals that depend higher than
quadratically on the states $|\vphi_k(T)\rangle$, linear equations of
motion and linear coupling to the control, it reads~\cite{ReichJCP12}
\begin{equation}
  \label{eq:sigma}
  \sigma(t) = -\bar{A}
\end{equation}
with $\bar{A}=\max\left(\varepsilon_A,2A+\varepsilon_A\right)$, where
$\varepsilon_A$ is a small non-negative number that can be used to
enforce strict inequality in the second order optimality condition.
The parameter $A$ depends on the final-time functional. In principle, it is
possible to determine a supremum for $A$ that guarantees convergence. In
practice, one should determine the optimal value for $A$ in each iteration
numerically~\cite{ReichJCP12},
\begin{widetext}
\begin{eqnarray}
  \label{eq:A_n}
  A^{(i+1)}  & =&
  \frac{\sum_{k=1}^{4}\left[
      \Braket{\chi_k\left(T\right)|\Delta\vphi_k\left(T\right)}
      +\Braket{\Delta\vphi_k\left(T\right)|\chi_k\left(T\right)}\right]
    +J_T\left(\{\varphi_k^{(i+1)}(T)\}\right)
    -J_T\left(\{\varphi_k^{(i)}(T)\}\right)}
  {\sum_{k=1}^{4}\left[
      \Braket{\Delta\vphi_k\left(T\right)|\Delta\vphi_k\left(T\right)}\right]}
  \,.
\end{eqnarray}
\end{widetext}

Evaluation of the update equation for the control,
Eq.~\eqref{eq:newu}, implies forward
propagation of the logical basis states and backward propagation of
the adjoint states. The forward propagation of the logical basis uses
the new control, as indicated by the superscript $(i+1)$,
\begin{subequations}\label{eq:forward_prop}
\begin{eqnarray}
  \frac{d}{dt}|\vphi^{(i+1)}_k(t)\rangle &=&
  -\frac{i}{\hbar}\Op{H}[u^{(i+1)}]
  |\vphi^{(i+1)}_k(t)\rangle \\
  |\vphi^{(i+1)}_k(0)\rangle &=& |k\rangle \,,\quad k=1,\ldots, 4   \,.
\end{eqnarray}
\end{subequations}
The adjoint states are propagated backward in time under the old
control, $u^{(i)}$,
\begin{subequations}\label{eq:backward_prop}
  \begin{eqnarray}\label{eq:chidot}
  \frac{d}{dt}|\chi^{(i)}_k(t)\rangle &=&
  -\frac{i}{\hbar}\Op{H}^{\dagger}[u^{(i)}]
  |\chi^{(i)}_k\rangle  \\
  |\chi^{(i)}_k(T)\rangle &=& -\nabla_{\Bra{\vphi_k}}
  J_T\big|_{|\vphi^{(i)}_k(T)\rangle}
  \, \label{eq:chiT}
  \end{eqnarray}
\end{subequations}
for the $k = 1 \dots 4$ states that constitute the logical two-qubit basis.
Note that it is Eq.~\eqref{eq:chiT} that necessitates a functional that is
differentiable with respect to the states.
The initial condition for the adjoint states is given in terms of the
final-time functional, $J_T$. We use either one of the functionals in
Eqs.~\eqref{eq:J_LI_tilde} and~\eqref{eq:J_pe_tilde}.

\section{Applications}
\label{sec:appl}

\subsection{Optimization for NV centers in diamond}
\label{subsec:NV}

\begin{figure*}[tb] 
  \centering
  \includegraphics{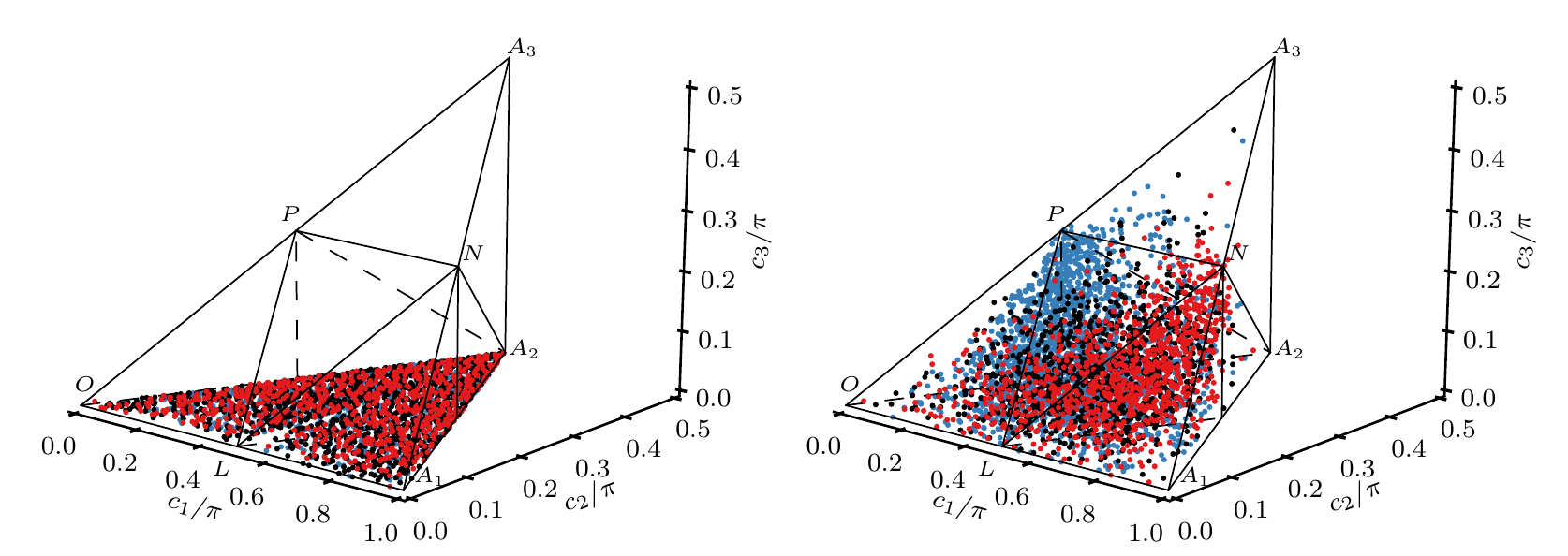}
  \caption{(Color online) Gates reached for an NV center in diamond
    with 2 control
    pulses (left) and 3 control pulses (right) during an optimization towards
    a perfect entangler (black), and towards the equivalence class of the
    points $P$ and $N$ in the Weyl chamber (blue and red, respectively).}
  \label{fig:NV_controllability}
\end{figure*}
For a system whose dynamics do not leak out of the logical subspace,
such as NV centers in diamond introduced in
Section~\ref{subsec:model:NV}, optimization towards
a perfect entangler is a powerful tool, given that half of all possible
two-qubit gates are perfect entanglers. Since without leakage any
control is guaranteed to yield a fully unitary gate, the likelihood of
finding a perfect entangler already with
an arbitrary control is large, provided the gate duration is sufficiently long.

In Fig.~\ref{fig:NV_controllability}, a sampling of all the gates obtained
during an optimization of the Hamiltonian~\eqref{eq:H_NV} is shown, for
two pulses, $\Omega_{MW}(t)$ and $\Omega_{RF(t)}$ (left),
cf.~Eqs.~\eqref{eq:H_MW} and~\eqref{eq:H_RF}, as well as for
a third control  $\Delta(t)$ (right), cf.~Eq.~\eqref{eq:DeltaControl}.
The gates from an optimization towards the polyhedron of perfect
entanglers, using the functional of Eq.~\eqref{eq:FUtilde-pE}, are indicated by
black dots.
The optimization was performed using the CRAB
algorithm and was allowed to continue even after reaching a perfect entangler.
Furthermore, an optimization towards the local equivalence class of the points
$P$ and $N$ (which are corners of the polyhedron of perfect entanglers -- see
Paper~I~\cite{Watts} for interpretation)
using the functional of Eq.~\eqref{eq:FUtilde-lec}, encountered
the gates shown by blue and red dots, respectively. In all cases, the results
confirm the predictions of Section~\ref{subsec:model:NV}: for two controls all
gates lie in the ground plane of the Weyl chamber, whereas for three controls
every part of the Weyl chamber is reached.

Since $P$ and $N$ are not reachable using only two control fields, the
optimization only yields success for the PE-functional, within at most
two iterations. Nonetheless, the
gates obtained from \emph{all} optimization targets (that is PE-functional and the
local equivalence classes P and N) sample the entire reachable
region; the black, red, and blue points in
Fig.~\ref{fig:NV_controllability} (left) each evenly cover the entire ground
plane of the Weyl chamber.

For three controls, the system shows full controllability, and the gates from
different optimization targets cluster in different regions. The gates from the
PE optimization evenly fill most of the front half of the PE polyhedron, whereas
the local-equivalence-class optimizations cluster in the direction of their
respective target points. In all cases, the desired target is reached. However,
there is a dramatic difference in the effort required in the different cases.
For the perfect entanglers, the optimization target was reached within usually
one or two optimization steps. In contrast, for the optimization towards the $P$
and $N$ points, at least several hundred iterations were necessary.

\subsection{Optimization for Josephson Charge Qubits}
\label{subsec:JCQ}

The optimization problem becomes more difficult once the model
accounts for the possibility of leakage out of the logical subspace,
which is the case for superconducting qubits. As a first example,
we optimize the system of coupled charge
qubits described in
Sec.~\ref{subsec:model:charge},  using the CRAB algorithm. For each qubit,
six levels were taken into account, and thus leakage from the logical subspace
had to be considered,
cf.~Eqs.~\eqref{eq:FUtilde-lec} and~\eqref{eq:FUtilde-pE}.
\begin{figure}[tb] 
  \centering
  \includegraphics{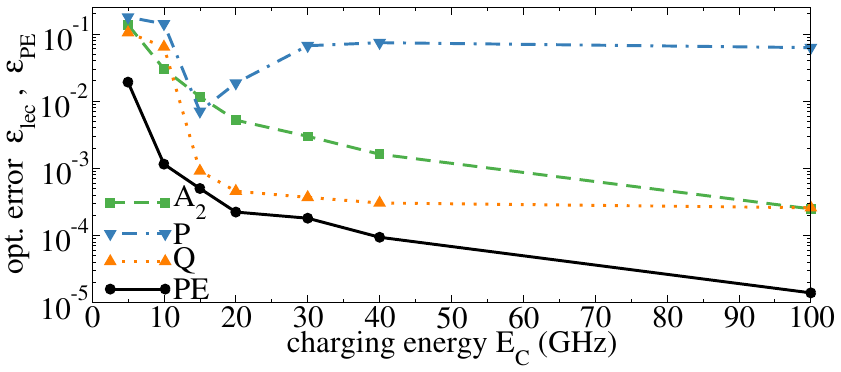}
  \caption{(Color online) Success of optimization for two Josephson
    charge qubits coupled by
  a Josephson junction. Control with one pulse $E_J(t)=E_{JJ}(t)$ and total time $T=1\,$ns.
 The error
  $\errPE$ for optimization toward the polyhedron of perfect
  entanglers (PE) is compared to the error $\errLec$
  for optimization towards the local equivalence classes of different
  points in the Weyl chamber $A_2$, $P$, and $Q$.}
  \label{fig:JCQ-fidelity-1pulse}
\end{figure}
\begin{figure}[tb] 
  \centering
  \includegraphics{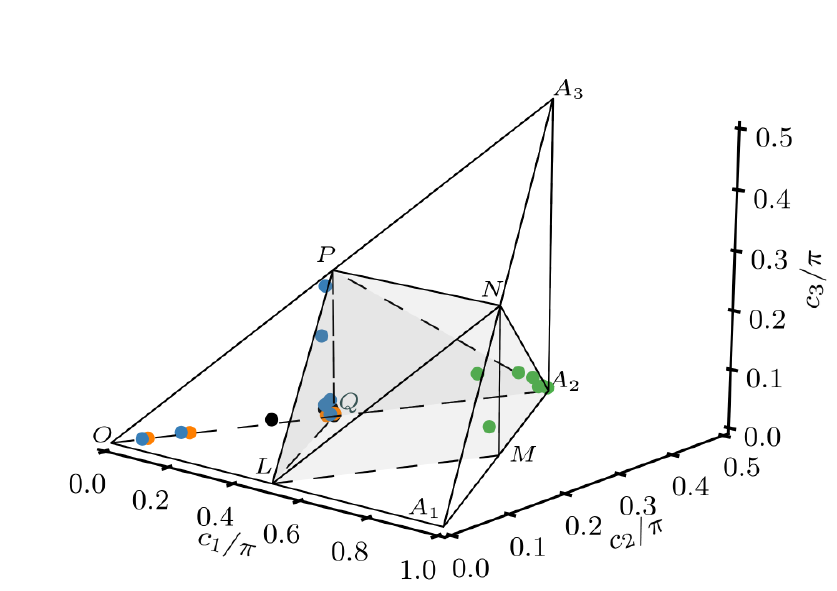}
  \caption{(Color online) Optimized gates in the Weyl chamber, for
    optimization of two coupled     charge qubits with
    $E_J(t)=E_{JJ}(t)$ and total time $T=1\,$ns,
    using the PE functional as well as the
    local invariants functional for the points $A$, $P$, $Q$.  The end
    points of the optimization for the results shown in
    Fig.~\ref{fig:JCQ-fidelity-1pulse}
    are indicated, using the same color coding (black for PE
    optimization, green for optimization towards the $A$ point, blue
    for $P$, and orange for $Q$).
  }
  \label{fig:JCQ-1pulse_weyl_paths}
\end{figure}
First, we consider the case $E_{JJ}(t)=E_J(t)$, i.e.\ using only a single
control pulse.
Figure~\ref{fig:JCQ-fidelity-1pulse} shows the optimization results of the
perfect entanglers functional for different values of $E_C$
and compares it to optimization towards a specific perfect
entangler equivalence class, for three corners of the polyhedron, $Q$, $A_2$,
and $P$. Success is measured by the error $\errPE(\tilde
U)=1-F_{\PE}(\tilde{U})$ for optimization towards the polyhedron of
perfect entanglers (black solid line, circles), and equivalently,
$\errLec(\tilde U)=1-F_{\lec}(\tilde{U})$ for optimization towards a
local equivalence class. Larger values of $E_C$ increase the spacing
between levels and thus make the implementation of a gate easier
as leakage of population to higher levels is suppressed.
We stress that while for the truncated Hamiltonian we can show full
controllability, this does not clearly imply full controllability also when
additional levels are included. Furthermore, the choice of specific parameters
can make certain parts of the Weyl chamber harder to reach in the chosen total time.
This is indeed what we see and report in Fig.~\ref{fig:JCQ-fidelity-1pulse}:
the optimization for an arbitrary perfect entangler significantly
outperforms the optimization towards a specific local equivalence
class.
Indeed, we find that while $Q$ and $A_2$ can be reached with
an error $\errLec < 10^{-3}$ and decreasing with the gate duration
(or equivalently increasing $E_C$), $P$ cannot be reached with
precision higher than a few percent.
This finding is supported by Fig.~\ref{fig:JCQ-1pulse_weyl_paths} that shows the
position in the Weyl chamber of the gates reached during optimization. Each dot
corresponds to a reached gate at final time (if leakage is present,
the closest gate $\in SU(4)$ is shown; see section~\ref{sec:CRAB}). This explains
the blue dot appearing near the point $P$: While the projection onto
$SU(4)$ gets relatively close to $P$, a loss of population from the
logical subspace of 2.1\% is observed, making the overall fidelity
low. In contrast, the population loss for the
gates ($Q$, $A_2$) as well as the
perfect entanglers is less then 0.1\%.
Interestingly, all optimizations towards a perfect entangler cluster
around the $Q$ point, which is  the
point that was reached with highest fidelity by a direct
local equivalence class optimization.

\begin{figure}[tb] 
  \centering
  \includegraphics{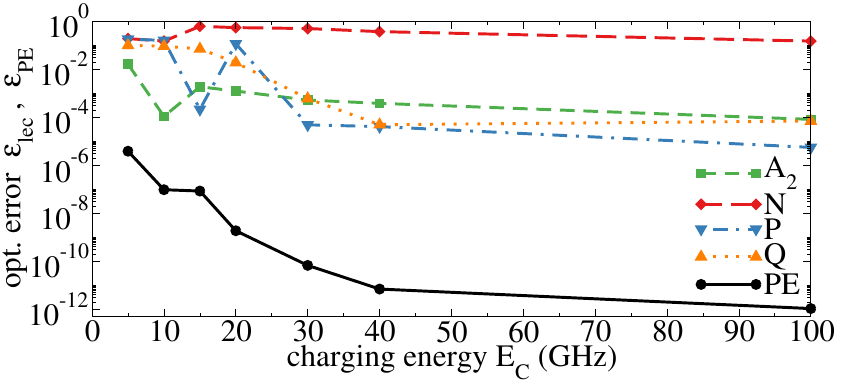}
  \caption{(Color online) Same as Fig.~\ref{fig:JCQ-fidelity-1pulse}
    but  control with two
    independent pulses $E_J(t)$ and $E_{JJ}(t)$.}
  \label{fig:JCQ-fidelity-2pulses}
\end{figure}
\begin{figure}[tb] 
  \centering
  \includegraphics{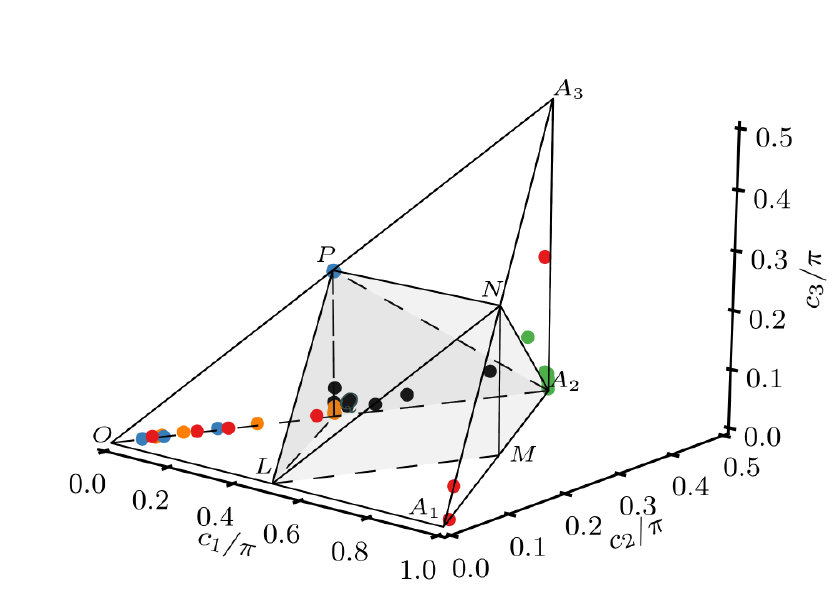}
  \caption{(Color online) Same as Fig.~\ref{fig:JCQ-1pulse_weyl_paths}
    but control with 2 pulses $E_J(t)$
    and $E_{JJ}(t)$. Color coding as in
    Fig.~\ref{fig:JCQ-fidelity-2pulses}.
  }
  \label{fig:JCQ-2pulses_weyl_paths}
\end{figure}
Finally, we relax the constraint $E_{JJ}(t)=E_J(t)$, allowing for two independent pulses:
Figure~\ref{fig:JCQ-fidelity-2pulses} shows the optimization success of the
perfect entanglers functional for different gate
durations, comparing it to optimization towards a given local
equivalence class. Here, we examine four corners of the
polyhedron, namely $Q$, $A_2$, $P$, and also $N$. As expected, again
the smallest errors, i.e., highest fidelities, are obtained
for perfect entangler optimization. In addition to $Q$ and $A_2$,
now also $P$ could be implemented with high fidelity, but $N$ (data
not included in Fig.~\ref{fig:JCQ-fidelity-1pulse}) remains unreachable. This is
also supported by Fig.~\ref{fig:JCQ-2pulses_weyl_paths} that shows the position
in the Weyl chamber of the gates reached during optimization. The obtained
results for two independent pulses suffer from significantly less loss of
population from the logical subspace compared to optimization with a single
pulse.
This is the expected behavior as the system goes
from being weakly controllable (the drift Hamiltonian is needed to
counteract leakage) for one control to being fully controllable for
two controls \cite{MotzoiPRL2009}.
For the optimization towards specific points in the Weyl chamber, the
loss was below 0.01\%, for the perfect entanglers it was as low as machine precision.

\subsection{Optimization of Transmon Qubits}
\label{subsec:SC}

Lastly, for two transmons as described in
Sec.~\ref{subsec:model:transmon},
we analyze the performance of the perfect entanglers functional using Krotov's
method, outlined in Sec.~\ref{sec:Krotov}. The optimization is carried
out for different gate durations between 25$\,$ns and 400$\,$ns,
starting from a $\sin$-squared  guess pulse of 35$\,$MHz peak amplitude.

\begin{figure}[tb] 
  \centering
  \includegraphics{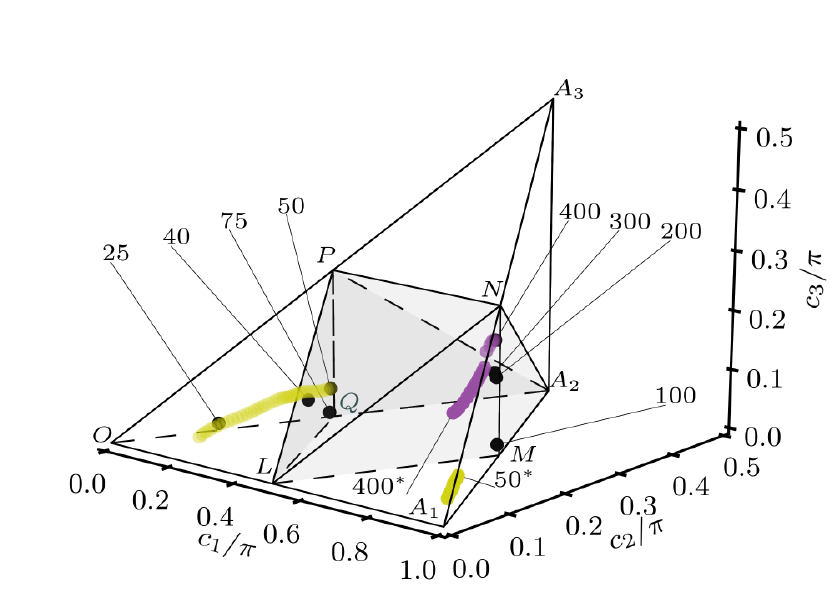}
  \caption{(Color online) Optimized gates in the Weyl chamber, for two
    transmon qubits, optimized with Krotov's method for the
    perfect entangler functional in Eq.~\eqref{eq:J_pe_tilde}. The
    point at which each optimization enters the PE polyhedron, or the
    end point of the optimization if no PE can be achieved,
    is shown by a black dot and labeled with the gate duration.
    The entire optimization paths for  $T=50\,$ns and $T=400\,$ns are
    shown in light blue and dark purple, respectively, with the
    starting points labeled by 50$^*$ and 400$^*$.}
  \label{fig:transmon_weyl_paths}
\end{figure}
Figure~\ref{fig:transmon_weyl_paths} shows the results of the
optimization in the Weyl chamber. The point at which each optimization
enters the perfect entanglers
polyhedron is indicated by a black dot and labeled
with the gate duration. For $T<50\,$ns, no perfect entangler can be
reached -- defining heuristically the QSL for this transformation.
In order to illustrate how the optimization proceeds,
the optimization paths for $T=50\,$ns, i.e., the gate at the
QSL, and a high-fidelity gate ($T=400\,$ns) are traced in light blue
and dark purple, respectively. Both optimizations start
in the $W_0^*$ region (near the $A_1$ point). The gate obtained with
the guess pulse for
$T=50\,$ns is significantly farther away from the surface of the
polyhedron of PE than that for the guess pulse with $T=400\,$ns.
Optimization for $T=400\,$ns therefore
moves directly towards the $W_0^*$ surface of the
PE polyhedron, whereas the optimization for $T=50$~ns enters the
ground plane and emerges in the $W_0$ region, before finally
reaching the $W_0$ surface of the polyhedron of perfect entanglers. The jump
from $W_0^*$ to $W_0$ is indicated by the light blue arrow. We find
the optimization to enter $W_0$ from $W_0^*$ for
durations $< 100\,$ns,  whereas for longer gate duration the optimizations
stay within $W_0^*$ entirely. The different optimization paths are a
result of the competition between the two objectives -- to reach a
perfect entangler, and to implement a gate that is unitary in the
logical subspace (the points shown in
Fig.~\ref{fig:transmon_weyl_paths} are the Weyl chamber
coordinates of the unitary $U$ closest to the actual time evolution
$\tilde{U}$). The latter objective is more difficult to realize
for shorter gate durations, resulting in a more indirect approach to
the polyhedron of perfect entanglers than one might expect when
considering that objective alone.

\begin{figure}[tb]
  \centering
  \includegraphics{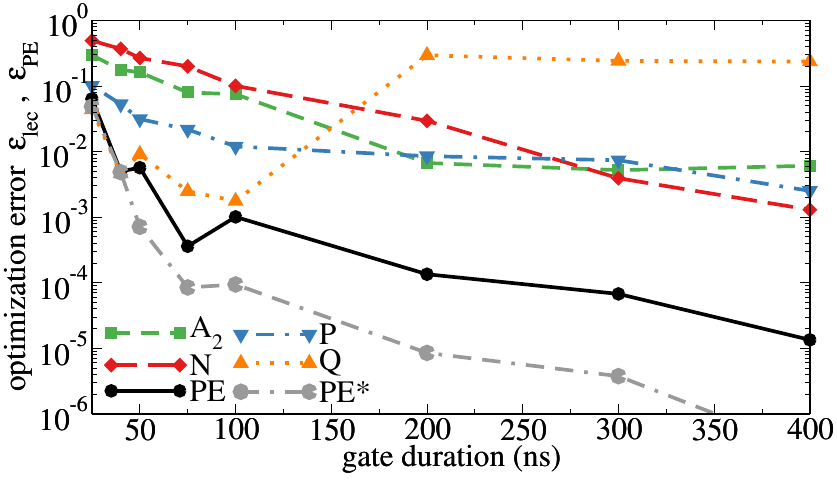}
  \caption{(Color online) Comparison of optimization success for the
    PE functional compared to the local invariants (LI) functional
    for several points in the Weyl chamber. The
    optimization success using Krotov's method
    is measured in $c$-space, although the optimization functionals are
    defined in $g$-space (see text for details). For the
    LI-optimization, the results are fully converged. For the PE-optimization,
    the results are converged to a relative change below
    $10^{-2}$ (black solid curve) and $10^{-3}$ (gray dash-dash-dotted curve).
  }
  \label{fig:tm_conv_LI}
\end{figure}
Analogously to the study of charge qubit optimization in
Sec.~\ref{subsec:JCQ}, it is instructive to compare the optimization
success of the perfect entangler functional,
Eq.~\eqref{eq:J_pe_tilde}, to that of the local invariants functional,
Eq.~\eqref{eq:J_LI_tilde}, for a few select points
of the Weyl chamber. This is shown in
Fig.~\ref{fig:tm_conv_LI}.
Instead of the optimization functionals Eqs.~\eqref{eq:J_pe_tilde} and
\eqref{eq:J_LI_tilde} we plot the gate errors $1-F_{\lec}(\tilde{U})$ for the
local invariants optimization, and $1-F_{\PE}(\tilde{U})$ for the PE
optimization, see. Eqs.~\eqref{eq:FUtilde-lec}
and~\eqref{eq:FUtilde-pE}.

The results of Fig.~\ref{fig:tm_conv_LI} show how, for different gate
durations, the gates that are easiest to reach differ. In agreement
with the results of Fig.~\ref{fig:transmon_weyl_paths}, for durations
$< 50$~ns, the jump in the optimization error indicates a speed limit.
For short gate durations, $50~\text{ns} \le T \le 100~\text{ns}$, optimization
towards the point $Q$ in the Weyl chamber is most successful. This
matches the optimized gates for $T \le 100\,$ns in
Fig.~\ref{fig:transmon_weyl_paths} being near the $Q$ point. Also
correspondingly, the longer gate durations end near the $N$ point. The
failure to reach the point $Q$ for longer durations is due to the
symmetry structure of the Weyl chamber.
Namely, for the ground plane of the chamber, there is a mirror axis defined
by the line through $L$ and $A_2$, where mirrored points are in the same local
equivalence class. Both the $Q$-point and the $M$ point have local
invariants of $g_1 = \frac{1}{4}, g_2=0, g_3=1$. Since the
optimization was performed in $g$-space, these two points are not
distinguishable; indeed, for long gate
durations, the $Q$-optimization successfully reached the $M$
point.

In comparison with the local invariants optimization, the perfect
entanglers functional shows excellent performance. It automatically identifies
the optimal gate for a given gate duration and reaches significantly better
fidelities. This is due to the fact that the desired entangling power of $U$ can
usually be obtained in just a few tens of iterations of the algorithm, and the
remainder of the optimization
then focuses on improving the unitarity of the obtained gate $\tilde{U}$.
Most strikingly, we find that for the optimization towards a specific local
equivalence class, the convergence rate becomes extremely small as the optimum
is approached. All the results shown in Fig.~\ref{fig:tm_conv_LI} are
converged to a relative change below $10^{-4}$, such that no measurable
improvement can be expected within a reasonable number of iterations.
While in principle (due to the full controllability
of the system), the direct optimizations should yield arbitrarily
small gate errors, as long as the gate duration is above the quantum
speed limit, in practice this depends
on numerical parameters such as the weight $\lambda_a$ in Krotov's
method and may take a extremely large number of iterations or stagnate,
as we observe here. The perfect
entangler optimization shows remarkable robustness with respect to
this issue. We observed very little slow-down in convergence. The black
curve in Fig.~\ref{fig:tm_conv_LI} for the PE-optimization already yields
a significantly smaller optimization error than any of the LI-optimizations, but
is only converged to a relative change of $10^{-2}$. Even the gray
dash-dash-dotted curve, labeled PE$^*$, is only converged to a relative change
of $10^{-3}$, and thus the optimization would still yield considerably better
results if it were to be continued.

\begin{figure}[tb] 
  \centering
  \includegraphics{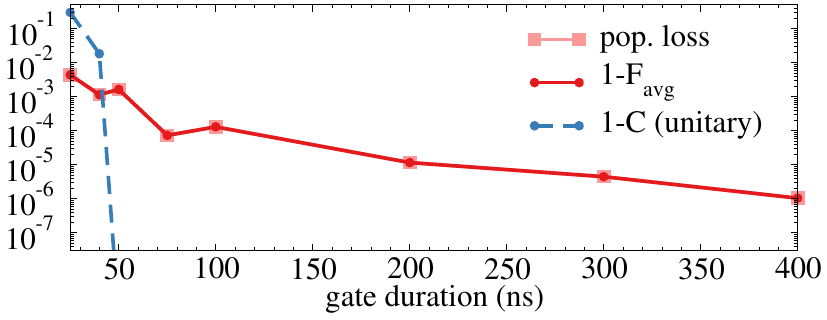}
  \caption{(Color online) Analysis of the sources of error for PE of
    2 transmon qubits:
    Population loss from the logical subspace (light red squares),
    concurrence error of the closest unitary gate $U$ in the logical
    subspace (blue circles), and average gate error,
    $\varepsilon_{avg}=1-F_{avg}$,  with which $U$ is
    implemented (red circles).
  }
  \label{fig:tm_conv_favg}
\end{figure}
The values of the optimization error in Fig.~\ref{fig:tm_conv_LI} of
$10^{-3}$ or $10^{-2}$
should not be understood to indicate a gate error above the quantum error
correction threshold.
Whereas the optimization error relates only to a figure of merit used for
optimization, the relevant physical quantity that would be determined in
an experiment is the
average gate fidelity.
It can be evaluated as~\cite{PedersenPLA2007}
\begin{eqnarray}
  F_{avg}
  & = &
    \frac{1}{20} \sum_{i,j=1}^4 \bigg(
              \langle \varphi_i |
                \Op{O}^\dagger
                \Op{U}\Ket{\varphi_i}\!\Bra{\varphi_j}\Op{U}^\dagger
                \Op{O} |
              \varphi_j \rangle
  \nonumber \\ & &
              + \Tr\left[
                \Op{O}\Ket{\varphi_i}\!\Bra{\varphi_i}\Op{O}^{\dagger}
                \Op{U}\Ket{\varphi_j}\!\Bra{\varphi_j}\Op{U}^\dagger
              \right]
           \bigg)\,,
\end{eqnarray}
assuming a two-qubit target gate $\Op O$ and denoting the logical
basis by $\{\Ket{\varphi_i}\}$.
Figure~\ref{fig:tm_conv_favg} shows the
generated entanglement as measured by the concurrence and the average
gate error, $\varepsilon_{avg}=1-F_{avg}$,
together with the population loss from the logical
subspace. $\Op O$ is taken to be the unitary that is closest
to the projection of the realized operation from the full Hilbert space
onto the logical subspace. For $T> 50\,$ns, the gate errors are at or
below $10^{-4}$.
For shorter gate durations, insufficient entanglement is generated,
cf. blue dashed curve in Fig.~\ref{fig:tm_conv_favg}. Once $T$ is
sufficiently large to generate the desired entanglement,
the only source of error is loss of population from the
logical subspace, shown in light red in
Fig.~\ref{fig:tm_conv_favg}. This loss does not depend on the choice
of the weight $w$ in Eq.~\eqref{eq:J_pe_tilde}. When the gate duration
is increased, optimization yields gates that are exponentially more
unitary, as indicated by the linear decrease of the average gate error
in our semi-log plot, in agreement with recently
introduced error bounds for optimal transformations~\cite{Lloyd2014}.
The difficulty to ensure unitarity on the logical subspace is typical
for weakly anharmonic ladders, as found in superconducting transmon or
phase qubits. Optimal control can be successfully employed to tackle
the problem of ensuring unitarity in the logical subspace,
in addition to generating entanglement, as exemplified in
Fig.~\ref{fig:tm_conv_favg}.

\section{Conclusions}
\label{sec:concl}

We have employed an optimization functional targeting an arbitrary
perfect entangler to obtain gate implementations for NV centers in
diamond and superconducting Josephson junctions. For NV centers in
diamonds, to a good approximation, the dynamics is confined to the
logical subspace. In this case, optimization for a
perfect entangler turns out to be trivial. This finding is in striking
contrast to optimization for a specific local equivalence class which
requires a large number of iterations, if it is successful at all.
The ease with which perfect entanglers are identified for
perfectly unitary time evolution can be rationalized by the fact that
more than
half of all non-local two-qubit gates are perfect
entanglers.

When population may leak out of the logical subspace, the optimization
problem becomes more difficult. Our corresponding examples were the
anharmonic ladders of superconducting qubits in the charge and
transmon architectures. While an optimization for a perfect entangler is
then no longer trivial, it converges much faster than an optimization
for a local equivalence class. This is rationalized by the larger
flexibility that a functional offers which allows for more possible
solutions. Larger flexibility implies an easier optimization problem, which is
reflected in better convergence properties of the algorithm, i.e.,
optimization is less likely to get stuck, and better final gate
fidelities can be reached. The perfect entanglers functional is
thus a better tool to investigate the quantum speed limit for perfectly entangling two-qubit gates,
i.e., the minimum time in which an perfectly entangling gate operation can be
performed, than the local invariants functional.

We find a qualitatively similar performance of two variants of the
perfect entanglers functional, expressed in terms of the Weyl chamber
coordinates and the local invariants. This is despite the very
different topologies connected with each formulation.

Our results underline the importance of properly expressing the
physical target in an optimization functional. This is particularly
encouraging in view of more complex quantum systems than those
considered here, including models that explicitly account for
decoherence. Such applications require identification of the unitary
that is closest to the actual dynamical map in order to evaluate the
perfect entanglers functional. This is possible by extending the
mathematical framework developed in
Refs.~\cite{GoerzNJP14,ReichPRA13} and will be subject of future
work.

\begin{acknowledgments}
We thank the Kavli Institute for
Theoretical Physics for hospitality and for supporting this 
research in part by the National Science Foundation Grant No. PHY11-25915.
We acknowledge the BWGrid for computational resources.
Financial support from the National Science Foundation under the
Catalzying International Collaborations program (Grant
No. OISE-1158954), the DAAD under grant PPP USA 54367416, the EC
through the FP7-People IEF Marie Curie action QOC4QIP Grant
No. PIEF-GA-2009-254174, EU-IP projects SIQS, DIADEMS and the
DFG SFB/TRR21 and the Science Foundation Ireland under Principal
Investigator Award 10/IN.1/I3013 is gratefully acknowledged.
\end{acknowledgments}


\end{document}